\documentclass[fleqn]{2023SCGE}
\usepackage{epstopdf} 

\begin{document}

\ensubject{subject}

\ArticleType{Article}
\SpecialTopic{SPECIAL TOPIC: }
\Year{***}
\Month{***}
\Vol{***}
\No{***}
\DOI{***}
\ArtNo{000000}
\ReceiveDate{***}
\AcceptDate{***}

\title{A Novel Fine Spectral Structure of Solar Radio Bursts with Periodic Beaded Stripes Observed by CBSm of CMP-II}

\author[1]{Chuanyang Li}{}%
\author[2,3]{Yao Chen}{yaochen@sdu.edu.cn}
\author[3]{Bing Wang}{}
\author[2]{Ze Zhong}{}
\author[4]{Baolin Tan}{}
\author[5]{Zongjun Ning}{}
\author[2]{\\Hao Ning}{}%
\author[2,3]{Xiangliang Kong}{}
\author[6]{Shuwang Chang}{}
\author[1]{Yanke Tang}{}
\author[1]{Ning Gai}{}
\author[7]{\\Li Deng}{}
\author[7]{Jingye Yan}{}
\author[3,6]{Fabao Yan}{}

\AuthorMark{Chuanyang Li}

\AuthorCitation{Chuanyang Li, Yao Chen, Bing Wang, et al}

\address[1]{College of Physics and Electronic Information, Dezhou University, Dezhou 253023, China}
\address[2]{Center for Integrated Research on Space Science, Astronomy, and Physics,\\ Institute of Frontier and Interdisciplinary Science, Shandong University, Qingdao 266237, China}
\address[3]{Institute of Space Sciences, Shandong University, Weihai 264209, China}
\address[4]{National Astronomical Observatories of Chinese Academy of Sciences, Beijing 100190, China}
\address[5]{Purple Mountain Observatory, Chinese Academy of Science, Nanjing 210008, China}
\address[6]{Laboratory for ElectronAgnetic Detection (LEAD), Institute of Space Science, Shandong University, Weihai 264209, China}
\address[7]{National Space Science Center, Chinese Academy of Sciences, Beijing 100190, China}


\abstract{A novel fine spectral structure in solar radio bursts has been discovered using the Chashan broadband solar radio spectrometer at meter wavelengths (CBSm), an instrument of the Chinese Meridian Project-Phase II (CMP-II). The structure features periodic narrow-band stripes with a typical recurrence time $< 1 $ s (occasionally reaches 8 s), often drifting from high to low frequencies and accompanied by absorptions, with trailing stripes appearing at the end of preceding ones. Some stripes exhibit periodic beaded enhancements with a periodicity of $\sim$0.1 s. The beaded stripes are reported for the first time ever. Data from the DAocheng Radio Telescope (DART) indicate a radio emission brightness temperature exceeding $10^{9}$ K, originating above brightening loops in active region AR 13664. We proposed a novel generation mechanism of the periodic stripes on the basis of the double plasma resonance (DPR) instability, and explained the beaded substructure in terms of modulation by low-frequency magnetohydrodynamic (MHD) waves. The study highlights the CBSm's capability to detect high-resolution fine spectral structures and offers novel insights into the emission mechanism and source characteristics of solar radio bursts.}



\maketitle


\begin{multicols}{2}
\section{Introduction}\label{Introduction}
Solar radio bursts, generated by coherent/incoherent emission processes from energetic electrons in the solar atmosphere, span a broad electromagnetic spectrum from decameter to millimeter wavelengths. \Authorfootnote
\noindent The study of these bursts plays a crucial role in understanding the physics underlying magnetic energy release, particle acceleration, and electromagnetic radiations in the solar atmosphere.

Type IV solar radio bursts, first classified by Boischot \cite{ref1}, are characterized by broadband continuum emission from the solar atmosphere at decimetric-decametric wavelength. Over several decades, various solar radio spectrograph systems worldwide, including Tremsdorf, IZMIRAN, Ondrejov, ARTEMIS, and CSO, have revealed numerous spectral fine structures in type-IVs (see \cite{ref2,ref3,ref4,ref5,ref6,ref7,ref8}). These include zebra patterns (ZPs), intermediate drift bursts (IDBs), radio spikes, fiber bursts (FBs), and slowly drifting chains of narrow-band fibers \cite{ref4,ref6} (also termed as packets of fibers \cite{ref5}, ropes of narrow-band fibers \cite{ref7}, or sawtooth bursts \cite{ref9,ref10}).

Earlier studies have revealed the key characteristics of narrow-band stripes: (1) a narrow bandwidth of $\sim$2 MHz and a slow frequency drift comparable to type II bursts, which is around 1 MHz/s. The drift can be either positive or negative \cite{ref2}; (2) the stripes are nearly parallel to each other sharing similar characteristics such as comparable drift rates ($\sim$-5 MHz/s), instantaneous bandwidth ($\sim$0.5 MHz), and range of frequency span. Typical fibers exhibit negative drift rates and accompanied by absorptions on the low-frequency side. The absorption features represent the decrease in intensity at frequencies lower than the stripes. Their bandwidths are comparable to and sometimes wider than those of the stripes; (3) occurrence time of stripes in a chain is $\sim$0.4 s, each chain contains 10 to several dozen stripes, lasting for several to 10 seconds \cite{ref5}.

Four scenarios have been proposed to explain the narrow-band stripes: (1) Mann et al. \cite{ref5} linked the phenomenon to periodic whistler excitations triggered by perturbations of the critical loss-cone angle of the magnetic mirror system. (2) Chernov \cite{ref7,ref11} associated the stripes with whistlers, and suggested periodic particle injection or repetitive bouncing between two termination shocks to account for the periodicity. (3) Klassen et al. \cite{ref9} argued the sawtooth structures are nonthermal plasma emission due to density fluctuations driven by current instabilities in flaring loops or current sheets. (4) Karlicky et al. \cite{ref10} attributed the sawtooth bursts to the modulation of the upper hybrid (UH) waves --- excited by the double plasma resonance (DPR) --- by other waves or density/magnetic oscillations.

The ground-based space environment monitoring network (Chinese Meridian Project-Phase II, CMP-II), a major national scientific infrastructure, was built and operated during the past five years. It deployed several solar radio observation systems, including the DAocheng Radio Telescope (DART, \cite{ref12}), MingantU SpEctral Radioheliograph (MUSER, \cite{ref13,ref14}), and the Chashan Broadband Solar radio spectrometer at Meter wavelengths (CBSm, \cite{ref15}).

Among them, CBSm has high frequency-temporal resolution and sensitivity (76.29 kHz, 0.84 ms, 1 sfu), making it a leading instrument for observing fine structures in solar radio spectra. It has detected tens of meter-wave bursts, including 57 type II bursts till February 3, 2025 and other bursts with various fine structures (\href{http://47.104.87.104/MWRS/RadioBurstEvent/typeII/typeIIburst_show.html}
{gallery of CSO type II bursts}). Data from CBSm have been used to study various types of solar radio bursts, such as type I and II bursts, quasi-periodic pulsations, etc. \cite{ref17,ref18, ref19, ref20}.

Upon examining CBSm spectral structures, we identified a novel spectral fine structure. They are closely associated with type-IV bursts and characterized by periodic narrow-band stripes, and some of which exhibit beaded enhancements --- termed as ``beaded stripes''.

This paper presents the first observation of beaded stripes with CBSm and investigates their formation mechanisms. The following section introduces the instruments and data. Section \ref{sec:3} presents the observational results, followed by a summary and discussion on the generation mechanism of the stripes.

\section{Instruments and Data}\label{sec:2}
The spectral data were obtained from CBSm in the range of 90--600 MHz. The CBSm system is located at the Chashan Solar Observatory (CSO) which was managed by the Institute of Space Sciences of Shandong University. It works with a 12 m parabolic reflector, a dual linear-polarized log-periodic feed, and a high-precision tracking platform. Its resolution is 76.29 kHz in frequency and 0.84 ms (up to 0.21 ms) in time. The dynamic range of the system, i.e., the level between the minimum discernible signal and the maximum undistorted signal, is $\sim$60--65 dB. Its sensitivity is $\sim$1 sfu for an integration time of 1 ms and a bandwidth of 100 kHz \cite{ref15}.

Additional data came from DART, which comprises 313 six-meter antennas uniformly distributed along a 1 km diameter ring. We obtained the DART images using the method of aperture synthesis. The circular array of DART produces a dish-shaped u-v coverage with $\sim$50000 quasi-uniform sampling of visibilities per polarization. The visibilities represent spatial-frequency Fourier components of the solar brightness distribution in space. Such dense and quasi-uniform sampling facilitates cleaner and high-fidelity reconstruction of solar images through direct Fourier inversion, without employing the computationally-expensive CLEAN algorithm. We used the original spectral or radio images given by CBSm and DART, without subtracting the background. With the aperture synthesis imaging technique across 16 frequency channels in range of 150--450 MHz (i.e., 149, 164, 190, 205, 223, 238, 285, 300, 309, 324, 366, 381, 399, 414, 432, and 447 MHz), DART achieves angular resolution of 1.5{$^{\prime}$} (450 MHz) to 5{$^{\prime}$} (149 MHz)\cite{ref12,ref22,ref23}.

We also analyzed the magnetogram data from the Helioseismic and Magnetic Imager (HMI: cadence: 45 s, pixel size: 0.5{$^{\prime\prime}$} \cite{ref24}) and the multi-wavelength extreme-ultraviolet (EUV) data from the Atmospheric Imaging Assembly (AIA: cadence: 12 s, pixel size: 0.6{$^{\prime\prime}$} \cite{ref25}), both onboard the Solar Dynamics Observatory (SDO: \cite{ref26}).

\section{Observational Results}\label{sec:3}
The event of interest occurred on May 8, 2024. We will first present an overview of this event before describing the novel spectral features.

\subsection{Event Overview}\label{sec:3.1}
\begin{figure}[H]
\centering
\includegraphics[width=0.95\columnwidth]{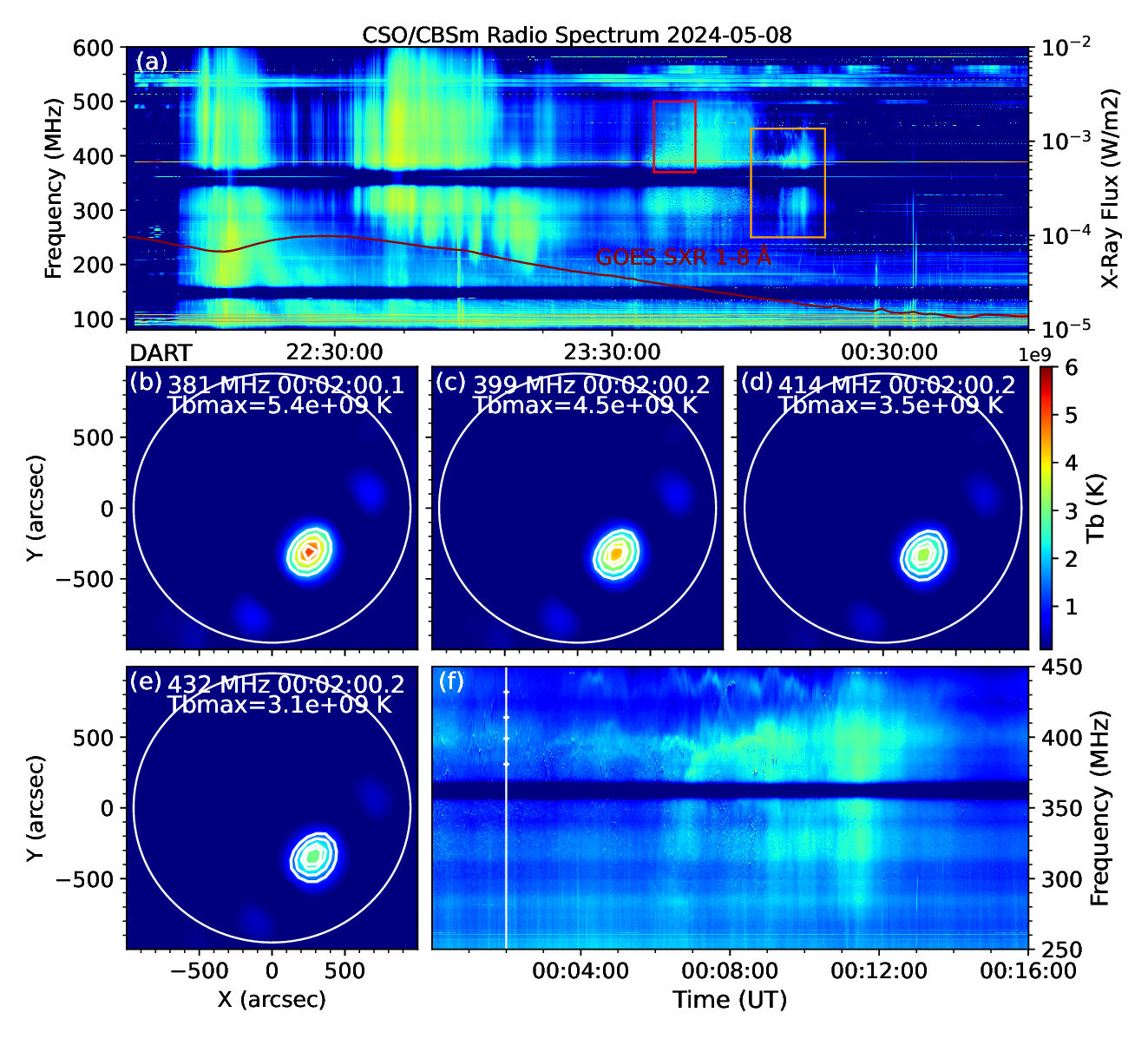}
\caption{Overview of the May 8, 2024 event. (a) Radio dynamic spectrum from CBSm (80--600 MHz, 22:00--01:00 UT) with overlaid GOES 1--8 \AA \ soft X-ray flux (deep red line). The orange box denotes the region displayed in panel (f), the red box shows the zoomed in region in Figure \ref{fig:3}. (b)--(e) DART radio source imaging at frequencies of 381 MHz, 399 MHz, 414 MHz, and 432 MHz, with contours
represent levels at 40\%, 60\%, 80\%, and 90\% of the maximum brightness temperature of respective frequency. (f) DART dynamic spectrum. The vertical white line denotes the time of the DART images, and the ticks mark the corresponding frequencies.
} 
\label{fig:1}
\end{figure}

Three radio enhancements of the event are shown in \cref{fig:1}(a). The first two correspond to X1.0 and M9.9 flares peaking at 21:40:00 UT and 22:27:00 UT, respectively. The third enhancement, occurring during the post-phase of the second flare, shows weaker intensity and narrower bandwidth (250--500 MHz) with rich fine structures.

\end{multicols}

\begin{multicols}{2}

\begin{figure}[H]
\centering
\includegraphics[width=0.85\columnwidth]{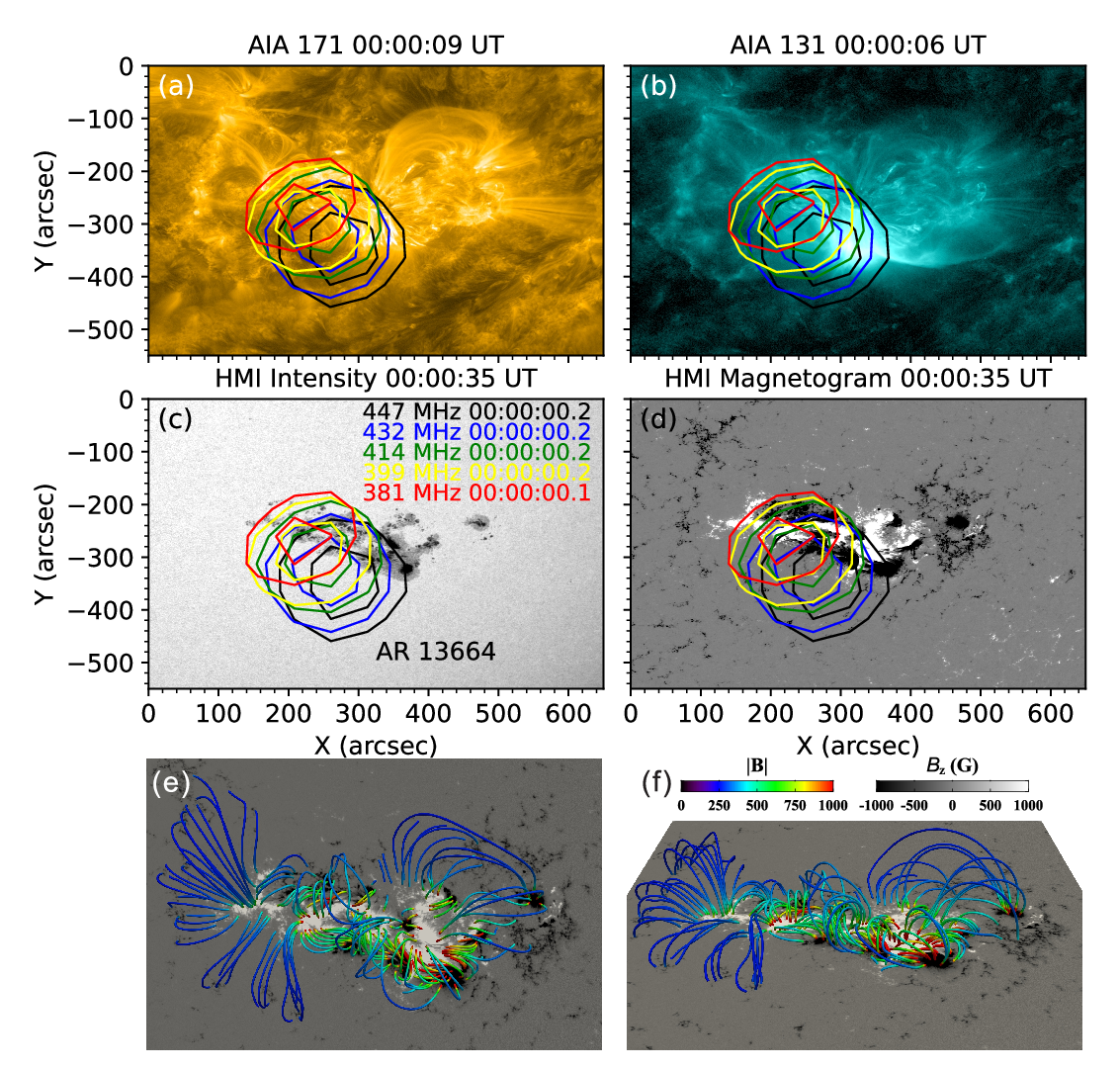}
\caption{(a) and (b) EUV images of AR13664 in AIA 171 \AA \ and 131 \AA. (c) and (d) HMI continuum intensity and line-of-sight magnetogram with overlaid contours of the simultaneous DART radio source. The contours correspond to imaging frequencies of 381 MHz, 399 MHz, 414 MHz, 432 MHz, and 447 MHz, with levels at 70\% and 90\% of the maximum brightness temperature of respective frequency. (e) and (f) NLFFF-extrapolated magnetic field lines.} 
\label{fig:2}
\end{figure}

DART reveals a bright source ($> 10^{9}$ K) southwest of the solar disk (Figures \ref{fig:1}(b)--(e)). Its dynamic spectrum in \cref{fig:1}(f) exhibits fine structures similar to those recorded by CBSm. The DART sources of the third radio enhancement are above the brightening loops southeast of AR 13664 (see Figures \ref{fig:2}(a) and (b)). HMI continuum and line-of-sight magnetogram (Figures \ref{fig:2}(c) and (d)) show complex magnetic topology with significant shear motions. According to the nonlinear force-free field (NLFFF) extrapolation (Figures \ref{fig:2}(e) and (f)), a complex system of loops exists underneath the DART sources.

\end{multicols}

\begin{multicols}{2}

\subsection{The Novel Spectral Fine Structure: Beaded Stripes}\label{sec:3.2}
\begin{figure}[H]
\centering
\includegraphics[width=0.85\columnwidth]{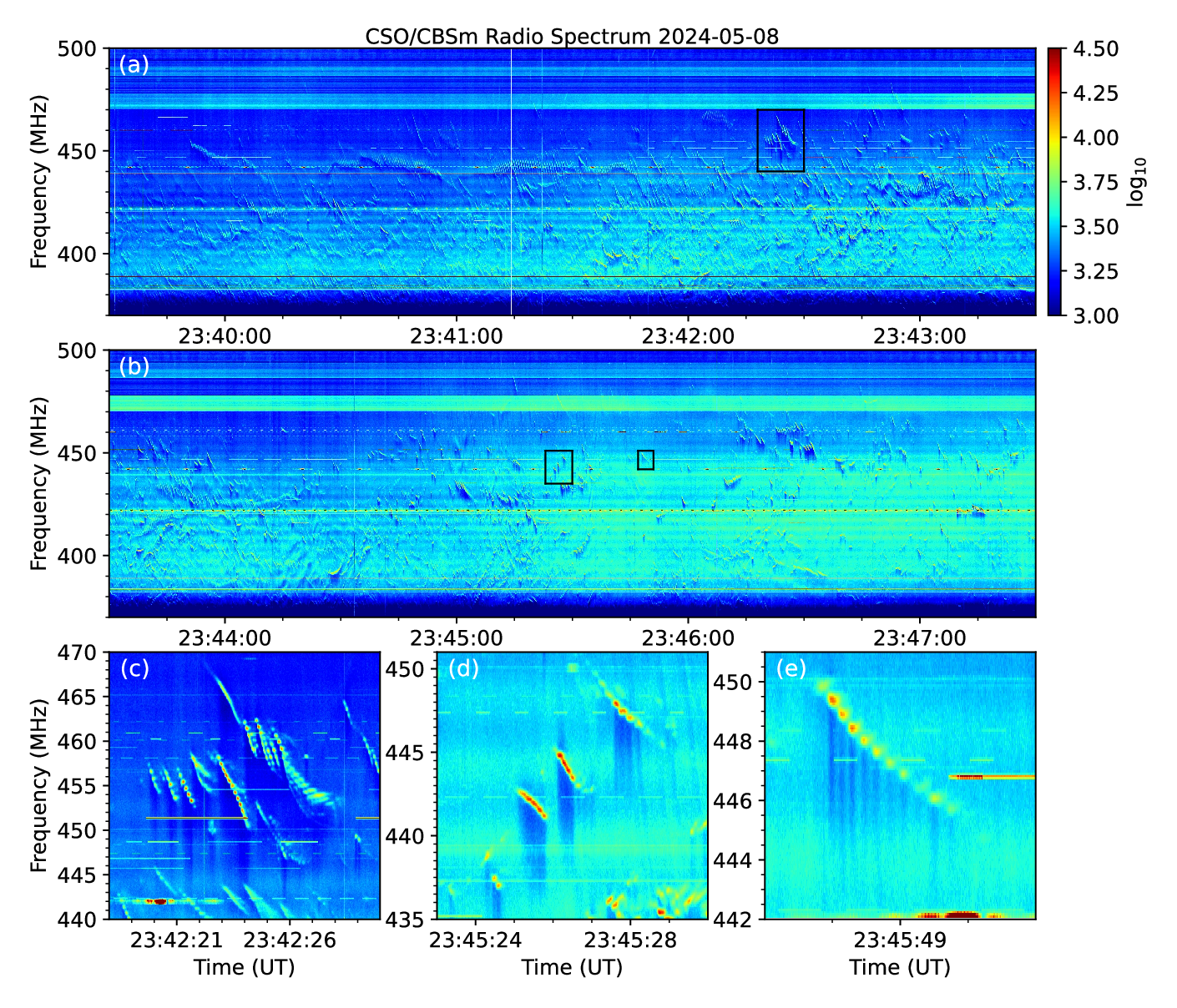}
\caption{(a) and (b) CBSm dynamic spectra containing rich fine structures (red box in \cref{fig:1}(a)). (c)--(e) Zoom-in views of the selected regions (black boxes in (a) and (b)).} 
\label{fig:3}
\end{figure}

Figures \ref{fig:3}(a) and (b) present zoomed-in dynamic spectra (370--500 MHz, 23:39:00--23:48:00 UT) corresponding to the red box in \cref{fig:1}(a), revealing abundant narrow-band stripes with diverse morphologies superimposed on the continuum emission. These stripes appear either individually or in a chain. Different chains/stripes may get overlapped with each other.

Figures \ref{fig:3}(c)--(e) reveal periodic enhancements of intensity along certain stripes, forming distinctive beaded pattern that has never been reported. Each stripe is associated with an absorption feature at its low-frequency side, with absorption bandwidths typically exceeding those of the emissions.

\end{multicols}

\begin{multicols}{2}

\begin{figure}[H]
\centering
\includegraphics[width=1\columnwidth]{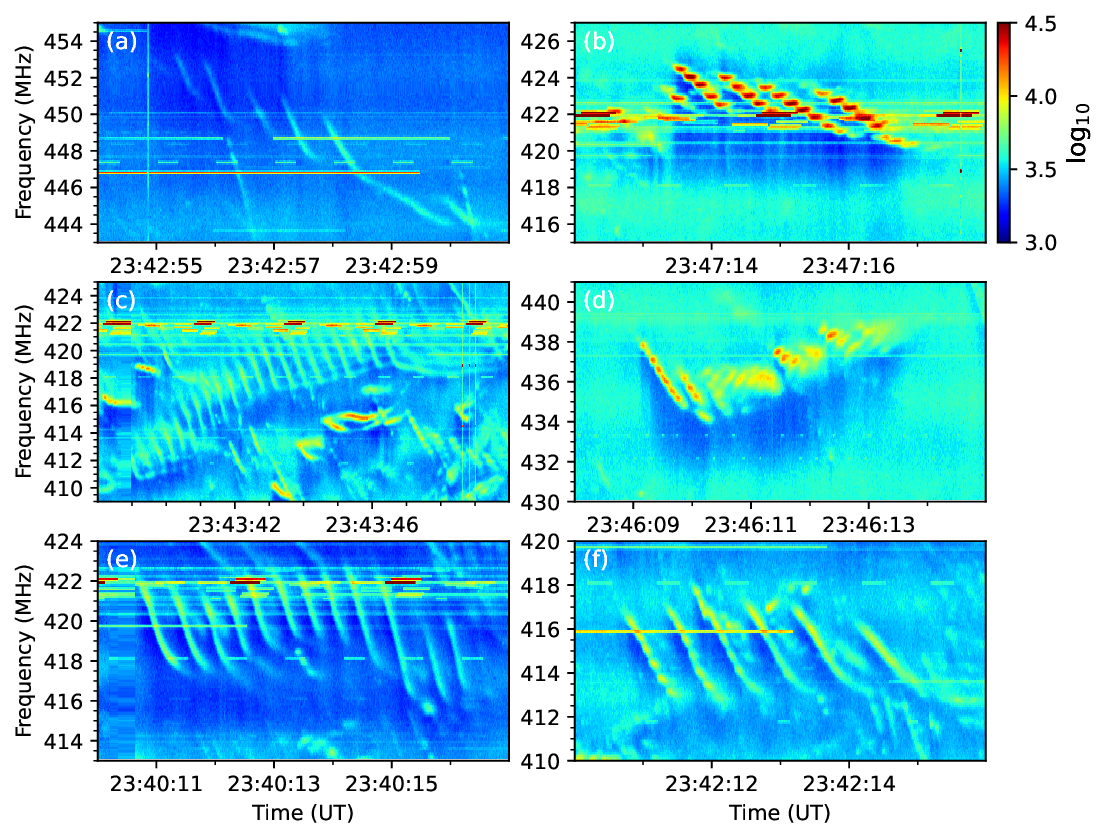}
\caption{Stripes with different drifting pattern in the May 8, 2024 event: negative drift (panels (a) and (b)), positive drift (panels (c) and (d)), and negligible drift (panels (e) and (f)). Left panels show stripes with no beads, right panels show beaded stripes.} 
\label{fig:4}
\end{figure}


The groups of stripes (chains) may have an overall negative (Figures \ref{fig:4}(a) and (b)), positive (Figures \ref{fig:4}(c) and (d)), or negligible (Figures \ref{fig:4}(e) and (f)) drift rate, while the individual stripes usually present negative drift rates. The stripes in a chain are basically parallel (i.e., with almost the same drift rates) to each other with remarkable periodicity ($\sim$0.5--0.8 s), yet with different start and end frequencies. The drift rates of stripes may vary by several MHz/s. For example, in Figure \ref{fig:4}(a), the drift rates of the first and last stripes vary from -12 MHz/s to -5 MHz/s, and in Figure \ref{fig:4}(e) the rates vary from -8 MHz/s to -13.5 MHz/s. Again, absorption features appear along the low-frequency edges.

\begin{figure}[H]
\centering
\includegraphics[width=0.81\columnwidth]{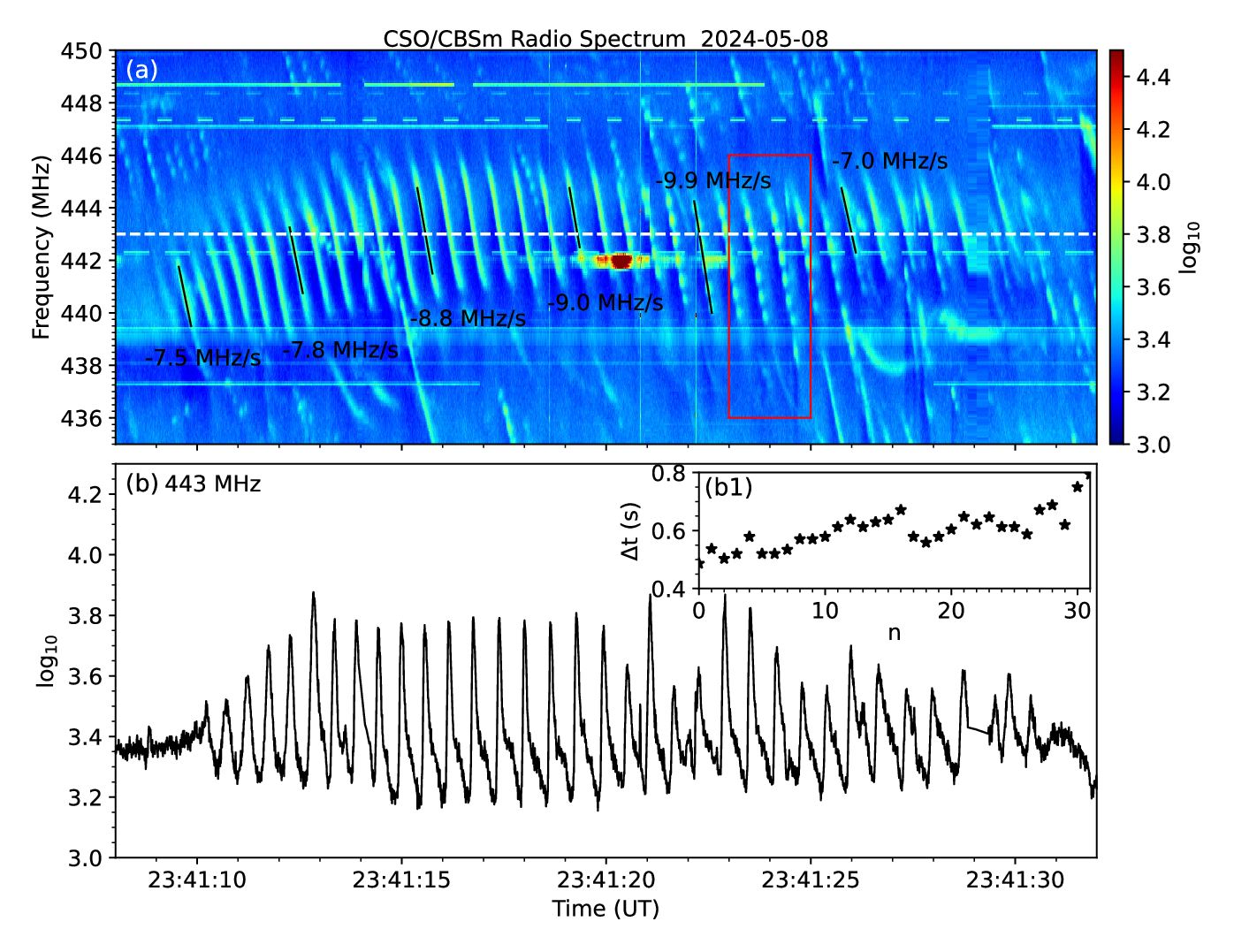}
\caption{(a) An example of stripes with an overall wavy morphology in the May 8, 2024 event, with beaded structure emerging in the middle portion. The black lines are fittings of the stripe slopes. The red box show the zoomed in view of \cref{fig:6}(a). (b) Intensity profile at 443 MHz (white dashed line in panel (a)). The inset (panel (b1)) shows the interval $\Delta$t between adjacent peaks.} 
\label{fig:5}
\end{figure}

\cref{fig:5} displays a chain with 35 stripes, according to which we observe: (1) a global wave-like drifting pattern: first drifting towards higher frequencies (23:41:10--23:41:15 UT), no-drifting (23:41:15--23:41:20 UT), towards lower frequencies until $\sim$23:41:25 UT, and then towards higher frequencies; (2) beaded structures emerge during the 6-minute interval (23:41:20--26 UT); (3) end-to-end temporal alignment, i.e., each stripe ends at the start of the next one.

\end{multicols}

\begin{figure}[H]
\centering
\includegraphics[width=0.95\columnwidth]{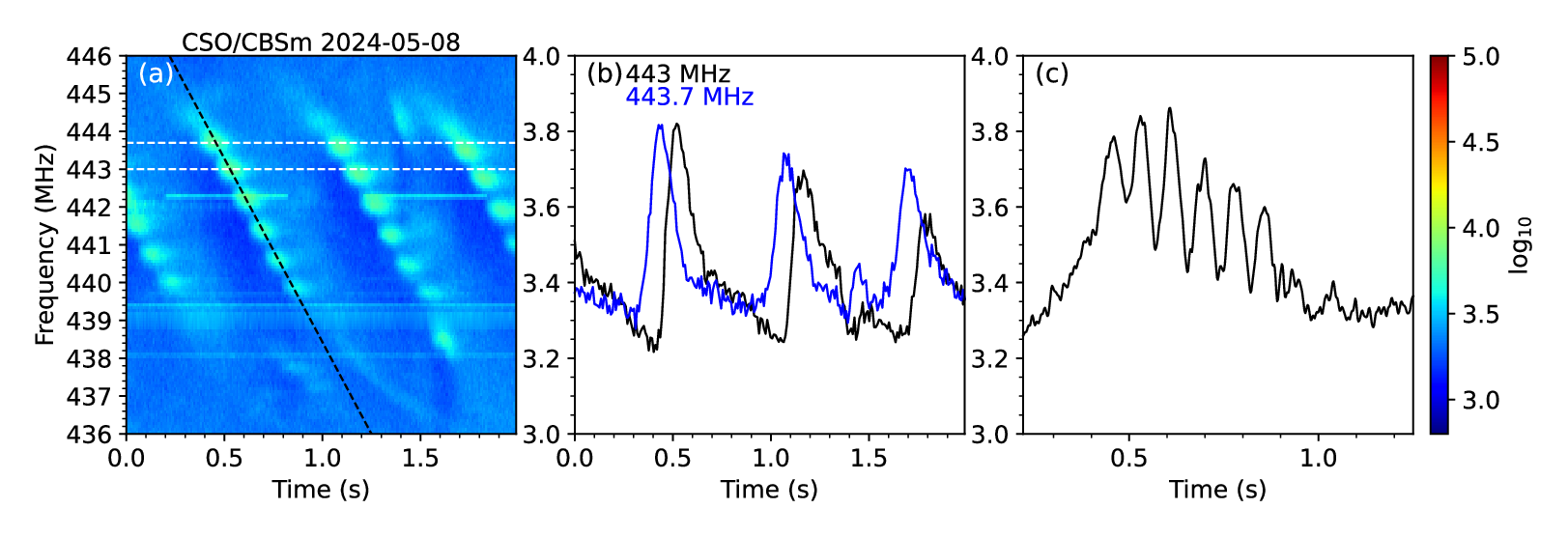}
\caption{ Zoom-in view of red box in \cref{fig:5}(a). (b) Temporal profiles at 443 MHz and 443.7 MHz along the white dashed lines plotted in panel (a). (c) Intensity variations along the black dashed line plotted in panel (a). } 
\label{fig:6}
\end{figure}

\begin{multicols}{2}

We measured the full-width-at-half-maximum (FWHM) of the stripe duration and the instantaneous bandwidth. The chain in \cref{fig:5} lasts $\sim$20 s, with individual stripes last 0.4--0.7 s, and their period is $\sim$0.5--0.8 s that gets longer gradually with time, according to the temporal profile at 443 MHz (see \cref{fig:5}(b)). The absorption manifests a intensity depression of $\sim$30\%, relative to the type IV continuum. The chain spans over 2--5 MHz with an instantaneous width of $\sim$1 MHz. The frequency drift rates vary from -7 to -10 MHz/s.

The parameters of the beads can be read from the zoom-in view of \cref{fig:6}, they are separated by $\sim$0.8 MHz in frequency and $\sim$0.1 s in time, their bandwidths are $\sim$0.6 MHz.

\subsection{Other CBSm Events with Chained Stripes}\label{sec:3.3}

\begin{figure}[H]
\centering
\includegraphics[width=0.79\columnwidth]{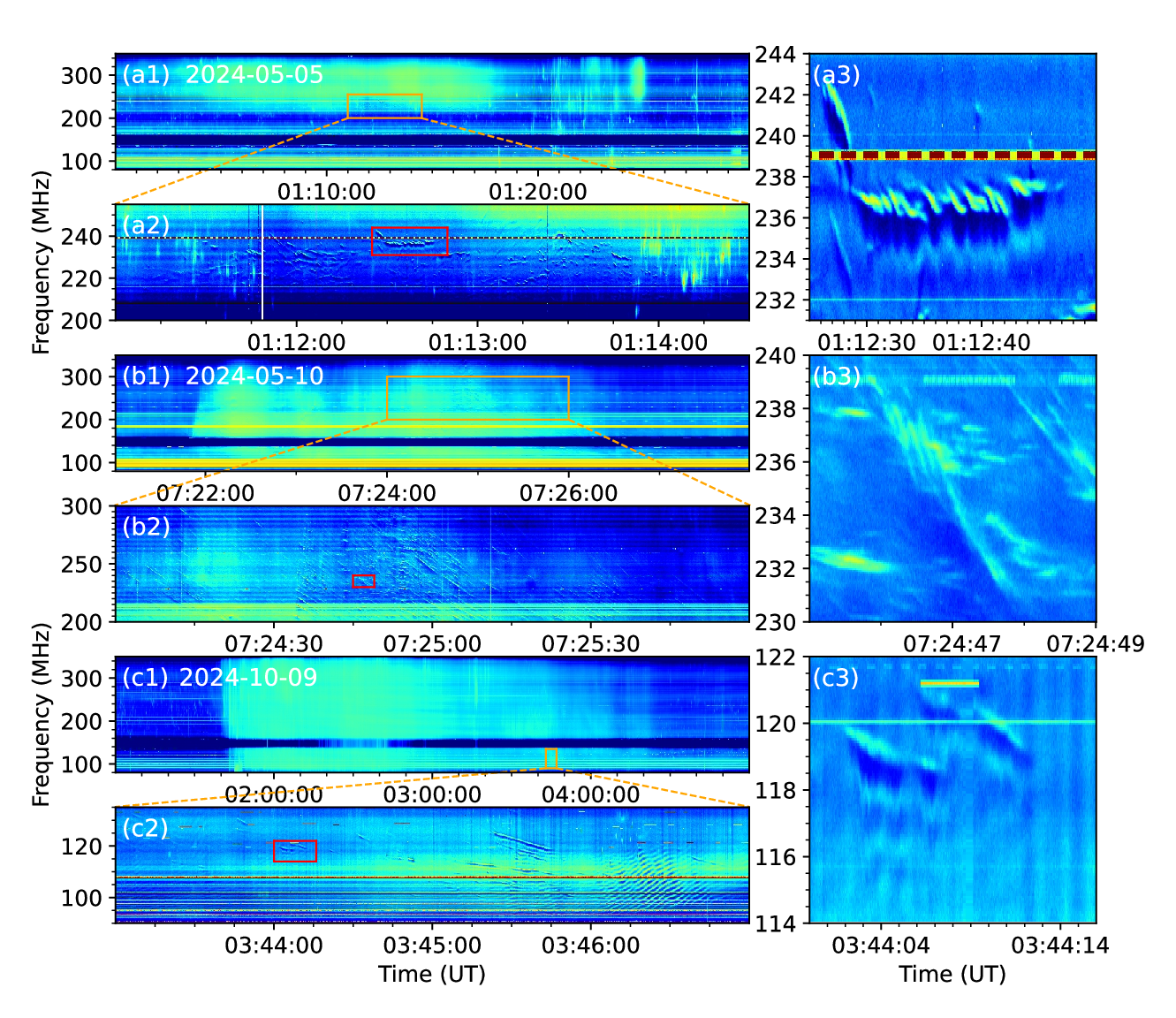}
\caption{Stripes observed by CBSm on May 5, 2024 (Event A, panels (a1)--(a3)), May 10, 2024 (Event B, panels (b1)--(b3)), and October 9, 2024 (Event C, panels (c1)--(c3)). The right panels are zoom-in views to observe details within the red boxes.} 
\label{fig:7}
\end{figure}

From May 5, 2024 to October 9, 2024, CBSm recorded three similar events with chained strips (see \cref{fig:7}). All present absorption features on the low-frequency side, yet without obvious beaded enhancements. In Table \ref{tab1}, we list the key characteristics of the four events.

\end{multicols}

\begin{multicols}{2}

Event A occurred in the pre-flare stage of the M8.4-class flare (peaking at $\sim$01:27:00 UT). It displays many narrowband stripes of emission and absorption. Figure \ref{fig:7}(a3) presents a typical group of stripes. The stripe period is $\sim$0.5--2 s with a narrower bandwidth ($\sim$0.5--1.5 MHz) and lower drift rates ($\sim$-1 -- -2 MHz/s) than the above event. Individual stripes last 0.5--1.5 s.

Event B occurred during the post-phase of the X3.9-class flare (peaking at $\sim$06:54:00 UT). It contains five stripes with bandwidths of $\sim$1.25--4 MHz and drift rates of $\sim$-6.5 -- -8 MHz/s, comparable to the May 8 event. The individual stripes last $\sim$0.2--0.6 s and the occurrence period is $\sim$0.1--0.2 s, shorter than the two events reported above.

Event C occurred during the post-phase of the X1.8-class flare (peaking at $\sim$01:56:00 UT). It has two stripes separated by $\sim$8 s. Each stripe has a bandwidth of $\sim$1--1.5 MHz, lasts about 2--5 s, and drifts at rates of $\sim$-0.25 MHz/s and -0.5 MHz/s, respectively.

\end{multicols}

\begin{table}[H]
\footnotesize
\begin{threeparttable}\caption{Key characteristics of the chained stripes of the four events}\label{tab1}
\doublerulesep 0.1pt \tabcolsep 13pt 
\begin{tabular}{ccccc}
\toprule
  Event                & Duration of stripes (s) & Period of stripes (s) & Bandwidth of stripes (MHz) & Drift rate (MHz/s) \\\hline
  2024.05.08           & 0.4--0.7 & 0.5--0.8 & 2--5 & -7 -- -10 \\
  2024.05.05 (Event A) & 0.5--1.5 & 0.5--2  & 0.5--1.5 & -1 -- -2  \\
  2024.05.10 (Event B) & 0.2--0.6 & 0.1--0.2 & 1.25--4 & -6.5 -- -8 \\
  2024.10.09 (Event C) & 2--5 & 8 & 1--1.5 & -0.25 -- -0.5 \\
\bottomrule
\end{tabular}
\end{threeparttable}
\end{table}

\begin{multicols}{2}

\section{Summary and Discussion}\label{sec:4}

Using the high-resolution spectral data of solar radio bursts recorded by CBSm of CMP-II, we identified a novel spectral structure of type-IV bursts with chained and beaded stripes. We observed four key characteristics: (1) each chain may contain several to dozens of stripes persisting for seconds to tens of seconds, most of them manifest negative drift rates of -1 to -10 MHz/s; (2) later stripes appear at the end of the former (in time), with the typical recurrence time $<1 $ s (occasionally it can be as high as 8 s) and bandwidth of 0.5--5 MHz (instantaneous bandwidth $\sim$1 MHz), and successive stripes tend to repeat in frequency but also can have different start/end frequencies; (3) the period of the beaded enhancements of stripes is $\sim$0.1 s in time and $\sim$0.8 MHz in frequency. (4) the low-frequency side of stripes is accompanied by features of absorption, and the relative intensity depression is $\sim$30\%.

Previous studies have reported chained stripes yet without the beaded substructures. Various mechanisms have been proposed to understand the generation of chained stripes. For instance, Karlicky et al. \cite{ref10} proposed a DPR-based scenario for similar sawtooth bursts, on the basis of the repetitive excitation of a single harmonic of the UH mode. According to our understanding, the stripes are due to trespassing periodic wave-like disturbances that carry the radiating source at a specific phase, and the wavelength should be about twice the source size to ensure end-to-end temporal alignment. It is difficult to explain why such disturbances exist.

Here we suggest a novel mechanism for such chains also on the basis of the DPR instability.

We suggest the radiations originate from UH waves via mode conversion. The UH waves are excited via the DPR instability that is driven by energetic electrons with unstable velocity distributions such as the loss-cone distribution. Such distribution excites UH waves at frequency close to $\omega_{UH}$ when $\omega_{UH} \approx s\Omega_{ce}$, in overdense plasmas with $\omega_{pe}/\Omega_{ce} \gg 1$ (e.g., refs. \cite{ref27,ref28,ref29,ref30}), where $s$ represents a positive integer.

\begin{figure}[H]
\centering
\includegraphics[width=0.95\columnwidth]{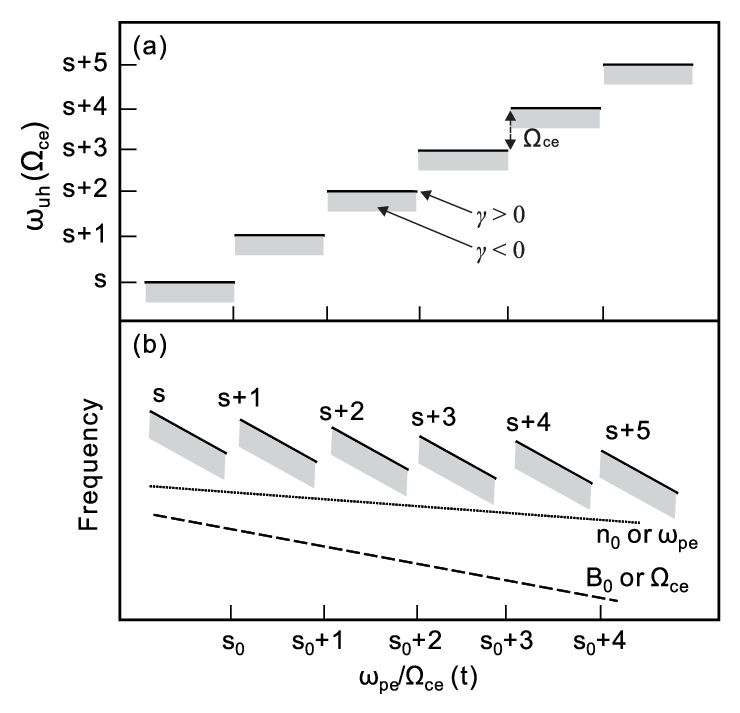}
\caption{Schematic diagram for the generation of striped chains. (a) Variations of the UH mode frequency as function of $\omega_{pe}/\Omega_{ce}$ (t). The shadows represent the absorptions, and $\gamma$ denotes the growth rate. (b) Schematic of chain formation due to the variations of $\omega_{pe}/\Omega_{ce}$ (t). } 
\label{fig:8}
\end{figure}

Temporal variations in plasma density ($n_0$) and/or magnetic field strength ($B_0$) can modulate the values of $\omega_{pe}/\Omega_{ce}$. Once $\omega_{pe}/\Omega_{ce}$ varies by a unity, i.e., from $s$ to $s+1$ or $s-1$, we have excitations of successive UH harmonics of ($s \pm 1$) $\Omega_{ce}$. The harmonics correspond to the radio stripes in a chain, as schemed in \cref{fig:8}. We propose that such variations (rather than a fixed UH harmonic mode) are the origin of periodic stripes. Since $\omega_{pe} \gg \Omega_{ce}$, the variation of $n_0$ determines the overall drift trend of the chain, while that of $B_0$ determines the drift of an individual stripe.

The absorption feature corresponds to the negative growth of UH waves. According to wave dispersion analysis (e.g., \cite{ref29}), efficient wave amplification can transform into strong wave damping with a small variation of frequency, and vice versa. This explains why such radiation-absorption pair occurs.

With this scenario we predict that (1) the drift of stripe chains mainly depends on $n_0$ and that of individual stripes mainly depends on $B_0$. This means one can evaluate such variations in the stripe source with the spectral data; (2) the frequency separation of successive stripes is determined by $\Omega_{ce}$, this allows us to diagnose $B_0$. For the four events reported here, we have $\Omega_{ce}$ varying from 1--5 MHz, then the magnetic field strength within the radio source varies from 0.4--1.8 G. This weak field may arise from the intersection of the multi-polar loop system overlying the sunpots (see Figures \ref{fig:2}(e) and (f)) where a null point may exist). Assuming the radiation to be the fundamental (harmonic) plasma emission, we have the plasma density within the burst source to be $2.5 \times 10^8$--$7.8 \times 10^8 \rm{cm}^{-3}$ ($1.9 \times 10^8$--$6.3 \times 10^8 \rm{cm}^{-3}$) for frequencies of 250--450 MHz.

We attribute the periodic beaded structures to modulation of the UH mode growth induced by the low-frequency MHD waves. The period of the beads is 0.1 s, this gives a wave frequency of $\sim$10 Hz. With the present data set, we cannot tell the exact mode, which might be one of the Alfv\'{e}n waves, fast/slow magnetoacoustic waves. In future, analysis of more events with similar feature and numerical simulation combining MHD wave dynamics and kinetic instabilities such as DPR are necessary for a better understanding of the wave growth and the modulated emission process.

In the following study, we will develop a quantitative model to evaluate the growth rates of the UH modes excited by the DPR instability, within temporally-evolving plasma density and/or magnetic field strength, so as to simulate the observed behavior of the chained stripes.

\section*{Data Availability Statement}

The DART data are available upon request from the DART team (yanjingye@nssc.ac.cn). The CBSm data can be accessed via the link: \href{http://47.104.87.104/MWRS/RadioBurstEvent/typeII/typeIIburst_show.html}{Gallery of CSO type-II Bursts}.

\Acknowledgements{This study is supported by by the National Natural Science Foundation of China (NNSFC) grants (Nos. 12103029, 11973031, 12303061, 12203031), the Natural Science Foundation of Shandong Province (NSFSP) grants (Nos. ZR2023QA074, ZR2021QA033), the National Key R\&D Program of China under grant 2022YFF0503002 (2022YFF0503000). We acknowledge the use of data from the Chinese Meridian Project. We thank the teams of CSO, DART, GOES, and SDO for making their data available to us. The authors are grateful to the anonymous referee for the valuable comments.}

\InterestConflict{The authors declare that they have no conflict of interest.}







\end{multicols}

\end{document}